\documentclass[12pt,twoside,a4paper]{article}
\usepackage{qmc,psfig}

\newcommand{\be}{\begin{equation}}
\newcommand{\ee}{\end{equation}}
\newcommand{\bea}{\begin{eqnarray}}
\newcommand{\eea}{\end{eqnarray}}

\begin{document}

\title{Thermodynamics from Monte Carlo Hamiltonian}

\author{ 
L.A. Caron$^{a}$ \\
H. Kr\"oger$^{a}$%
\footnote{talk given by H. Kr\"oger at QMC Trento 2001, email: hkroger@phy.ulaval.ca} \\
G. Melkonyan$^{a}$ \\
X.Q. Luo$^{b}$ \\
K.J.M. Moriarty$^{c}$
 }
\instit{
$^{a}$D\'epartement de Physique, Universit\'e Laval, \\
Qu\'ebec, Qu\'ebec G1K 7P4, Canada \\
$^{b}$Department of Physics, Zhongshan University, \\
Guangzhou 510275, China \\
$^{c}$Department of Mathematics, Statistics and Computer Science, \\
Dalhousie University, Halifax, Nova Scotia B3H 3J5, Canada
 }
\gdef\theauthor{ 
L.A. Caron, H. Kr\"oger, G. Melkonyan, X.Q. Luo, K. Moriarty
}
\gdef\thetitle{
Thermodynamics from Monte Carlo Hamiltonian
}

\maketitle

\begin{abstract}
We construct an effective low-energy Hamiltonian from the classical action via Monte Carlo with importance sampling. We use Monte Carlo (i) to compute matrix elements of the transition amplitude and (ii) to construct stochastically a basis. The MC Hamiltonian allows to obtain energies and wave functions 
of low-lying states. It allows also to compute thermodynamical observables 
in some temperature window (starting from temperature zero).  
We present examples from lattice field theory (Klein-Gordon model).
\end{abstract}

\section{Introduction}
\label{sec:Introduction}

The most interesting phenomena in physics occur in many-body systems.
The difficulty lies in solving models of such systems.
A brilliant idea to do this is the renormalisation group approach 
\`a la Kadanoff, Wilson, Migdal and others. It suggests to compute 
critical phenomena  
by constructing an effective (renormalized) Hamiltonian, by thinning 
out degrees of freedom in an appropriate manner.
Here we suggest to adapt this idea for the pupose to construct 
an effective Hamiltonian in a low-energy window.
The basic idea to thin out degrees of freedom is taken over from 
lattice field theory, where infinite-dimensional integrals 
(path integrals) are successfully simulated by a small 
number of representative configurations.
In analogy to the representative configurations of path integrals,
we construct a representative basis of Hilbert states 
via a Monte Carlo algorithm with importance sampling.  
Furthermore we use the Monte Carlo algorithm to compute matrix 
elements of the
transition amplitude.

In standard Lagrangian lattice field theory, one is used to construct a corresponding lattice Hamiltonian via the transfer matrix, which 
corresponds to
the transition amplitude for infinitesimal short time (when the lattice 
spacing $a_{t}$ goes to zero). As a result, one finds a Hamilton operator 
(on the lattice) which still has the character of a many-body Hamiltonian.
In contrast to that, we start out from the transition amplitude for a 
finite time (in imaginary time corresponding to some finite temperature).
We compute the transition matrix for all transitions, where the states 
are taken from the finite stochastically generated basis. As a result, 
we find an effective Hamiltonian, i.e. its spectrum and eigen states. 
It is evident that such Hamiltonian covers only a window of physics, 
which in this case is a temperature window, starting from temperature 
zero.   

From the viewpoint of the enormous success of standard Lagrangian 
lattice field theory, one may ask: Do we need such an effective 
Hamiltonian?
To give an example:
Spectrum and wave functions of excited states. Wave functions in 
conjunction with the energy spectrum contain more physical information 
than the energy spectrum alone. Although lattice $QCD$ simulations 
in the Lagrangian formulation
give good estimates of the hadron masses, one is yet far from a 
comprehensive understanding of hadrons. Let us take as example a 
new type of hardrons made of
gluons, the so-called glueballs. Lattice QCD calculations 
\cite{Luo:96}
predict the mass of the lightest glueball with quantum number 
$J^{PC}=0^{++}$,
to be $1650 \pm 100 MeV$. Experimentally, there are at least two
candidates: $f_0(1500)$ and $f_J(1710)$. The investigation of the 
glueball production and decays can certainly provide additional 
important information for
experimental determination of a glueball. Therefore, it is important 
to be able to compute the glueball wave function. 
The standard Lagrangian lattice formulation
has made slow progress also in some other areas, like on
hadron structure functions (not only moments) in particular at 
small $Q^{2}$ and at small $x_{B}$, as well as on scattering 
amplitudes.

\section{Effective Hamiltonian}
\label{sec:Effective Hamiltonian}

Let us discuss the construction of the effective Hamiltonian in 
several steps \cite{Jirari:99}. First, consider in quantum mechanics in 1-dim the motion 
of a single particle of mass $m$ under the influence of a local potential 
$V(x)$. Its classical action is given by
\be
S = \int dt ~ \frac{m}{2} \dot{x}^{2} - V(x) ~ .
\ee
Given the classical action, one can determine Q.M. transition amplitudes.
Like in Lagrangian lattice field theory, we use imaginary time in what follows. 
We consider the transition amplitude for some finite time $T$ ($t_{in}=0$, $t_{fi}=T$) and for all combinations of positions $x_{i}, x_{j}$.
Here $x_{i}$, $x_{j}$ run over a finite discrete set of points $\{x_{1},\dots,x_{N}\}$ located on the real axis. Suppose these points are equidistantly distributed (spacing $\Delta x$) over some interval $[-L,+L]$. 
The transition amplitudes are given by the (Euclidean) path integral,
\begin{equation}
\label{eq:TransAmpl}
M_{ij}(T)
= \int [dx] \exp[ - S_{E}[x]/\hbar ]\bigg |_{x_{j},0}^{x_{i},T} ,              
\end{equation}
where $S_{E}$ denotes the Euclidean action.
Suppose for the moment that those transition amplitudes were known. 
From that one can construct an approximate, i.e. effective Hamiltonian.
Note that the matrix $M(T)=[M_{ij}(T)]_{N \times N}$ is a real, positive and Hermitian matrix. It can be factorized into a unitary matrix $U$ 
and a real diagonal matrix $D(T)$, 
\be
\label{eq:Factor}
M(T)=U^{\dagger}~D(T)~U ~ .
\ee
Because the matrix $M(T)$ can be expressed in terms of the full Hamiltonian $H$
by
\be
M_{ij}(T) = \langle x_{i} | e^{-H T /\hbar} | x_{j} \rangle ~ ,
\ee
the matices $U$ and $D$ have the following meaning,
\bea
\label{eq:UDInterpret}
&& U^{\dagger}_{ik}=<x_i|E_k^{eff}>
\nonumber \\
&& D_k(T)=e^{-{E_k^{eff}}T/\hbar} ~ ,
\eea
i.e., the $k-th$ eigenvector $|E_k^{eff}>$ of the effective Hamiltonian $H_{eff}$ can be identified with the $k-th$ column of matrix $U^{\dagger}$ and
the energy eigenvalues $E^{eff}_{k}$ of $H_{eff}$ can be identified with the logarithm of the diagonal matrix elements of $D(T)$.
This yields an effective Hamiltonian, 
\be
H_{eff} = \sum_{k =1}^{N} | E^{eff}_{k} > E^{eff}_{k} < E^{eff}_{k} |.
\ee
Note that in the above we have been mathematically a bit sloppy.
The states $| x_{i} \rangle$ are not Hilbert states. We have to replace  
$| x_{i} \rangle$ by some "localized" Hilbert state. This can be done by introducing box states. We associate to each $x_{i}$ some box state $b_{i}$,
defined by
\be
b_{i}(x) =  \left\{ 
\begin{array}{l}
1/\sqrt{\Delta x_{i}} ~~ \mbox{if} ~~ x_{i} < x \leq x_{i+1}
\\
0 ~~ \mbox{else}
\end{array}
\right.
\ee
where $\Delta x_{i} = x_{i+1}-x_{i}$.
When we use an equidistant distribution of $x_{i}$, i.e. 
$\Delta x_{i} = \Delta x$, we refer to the basis of box states as the regular basis. Below we will consider a stochastic basis.

\section{Matrix elements}
\label{sec:MatElem}

We compute the matrix element $M_{ij}(T)$ directly from 
the action via Monte Carlo with importance sampling. 
This is done by writing each transition matrix element 
as a ratio of two path integrals.
This is done by splitting the action into a non-interacting part and an interacting part,
\be
S = S_{0} + S_{V} ~ .
\ee
This allows to express the transition amplitude by
\be
M_{ij}(T) = M^{(0)}_{ij}(T) ~
\frac{ 
\left.
\int_{x_{i}}^{x_{i+1}} d y 
\int_{x_{j}}^{x_{j+1}} d z
\int [dx] ~ \exp[ - S_{V}[x]/\hbar ] ~ \exp[ -S_{0}[x]/\hbar ] \right|^{y,T}_{z,0} }
{ \left.
\int_{x_{i}}^{x_{i+1}} d y 
\int_{x_{j}}^{x_{j+1}} d z
\int [dx] ~ \exp[ -S_{0}[x]/\hbar ] \right|^{y,T}_{z,0} } ,
\ee
Here $e^{-S_{0}/\hbar}$ is the weight factor and $e^{-S_{V}/\hbar}$ is the observable.
$M^{(0)}_{ij}(T)$ stands for the transition amplitude of the noninteracting system, which is (almost) known analytically.
For details see ref.\cite{Jirari:99}.
Carrying out these steps allows to construct an effective Hamiltonian, which reproduces well low energy physics, provided that the box functions cover an interval large enough (depending on the range of the potential) and the resolution $\Delta x$ is small enough. This can be achieved with a small numerical effort for a one-body system in 1 dimension.
Our goal is, however, to solve many-body systems.
For such purpose, the above regular basis fails. 
What to do in such systems is the subject of the following section.

\section{Stochastic basis}
\label{sec:Stochastic}

It is evident that the regular basis defined above becomes prohibitively large
when applied to a many-body system. For example, in a spin model of a 1-dimensional chain of 30 atoms with spin 1/2, the Hilbert space has the dimension $D=2^{30} = 1 073 741 824$.
For such situations we wish to construct a smaller basis which gives an effective Hamiltonian reproducing well low-energy observables.
Why should such a basis exist in the first place?
The heuristic argument is the Euclidean path integral, which, when evaluated 
via Monte Carlo with importance sampling, gives a good answer for the 
transition amplitude. In particular, this is possible by taking into account a "small" number of configurations (e.g. in the order of 100 - 1000). In a crude way the configurations correspond to basis functions. Thus we expect that suitably chosen basis functions exist, the number of which is in the order of 100 - 1000, which yields a satisfactory effective low energy Hamiltonian.

How can we construct such a "small" basis? As example consider a free particle in $D=1$ dimension. Recall: For the free system the transition amplitude reads
\be
G_{Eucl}(x,T;y,0) = \sqrt{ \frac{m}{2 \pi \hbar T} } 
\exp[ - \frac{m}{2 \hbar T} (x - y)^{2} ] .
\ee
This is a positive function for all $x$, $y$, $T$. It can be interpreted as a probability density.
We put $y=0$ and define a probability density $P(x)$ by
\bea
\label{eq:Probability}
P(x) &=& \frac{1}{Z} G_{Eucl}(x,T;0,0),
\nonumber \\
Z &=& \int dx ~ G_{Eucl}(x,T;0,0) .
\eea
Then we define a selection process as follows: Using a random process 
with probability density $P(x)$ one draws a "small" set of samples  
$\{ x_{\nu}| \nu \in 1, \dots, N_{eff} \}$.
In the case of the free particle, $P(x)$ is a Gaussian,
\be
\label{eq:Gaussian}
P(x) = \frac{1}{\sqrt{2 \pi} \sigma} \exp [ - \frac{x^{2}}{2 \sigma^{2}} ], ~~
\sigma = \sqrt{ \frac{\hbar T}{m} } .
\ee
In other words, we select $\{x_{\nu}\}$ by drawing from a Gaussian distribution. We draw $N_{eff}$ samples, giving $N_{eff}$ states, where $N_{eff}$ is considerably smaller than $N$, the original size of the basis.

Next we ask: What do we do in the case when a local potential is present? The definition of the probability density $P(x)$ given by Eq.(\ref{eq:Probability})
generalizes to include also local potentials.
In order to construct a stochastic basis one can proceed via the following steps:
(i) Compute the Euclidean Green's function $G_{E}(x,t;0,0)$, e.g., by solving the diffusion equation and compute $P(x)$. (ii) Find an algorithm giving a random variable $x$ distributed according to the probability density $P(x)$ and draw samples from this distribution, giving nodes, say $x_{\nu}$. Finally, one obtains the stochastic basis by constructing the corresponding characteristic states from the nodes $x_{\nu}$.  
The same goal can be achieved in an elegant and efficient manner via 
the Euclidean path integral. Writing Eq.(\ref{eq:Probability}) as path integral yields
\bea
\label{eq:ProbDensPathInt}
P(x) = \frac{ \int [dy] \exp[ - S_{E}[y]/\hbar ]\bigg|_{0,0}^{x,T} }
{ \int_{-\infty}^{+\infty} dx 
\int [dy] \exp[ - S_{E}[y]/\hbar ]\bigg|_{0,0}^{x,T} } ~~~ .
\eea
Using a Monte Carlo algorithm with importance sampling (e.g., Metropolis) one generates representative paths, which all start at $x=0$, $t=0$ and arrive 
at some position $x$ at time $t=T$.
Let us denote those paths (configurations) by $C_{\nu} \equiv x_{\nu}(t)$.
We denote the endpoint of path $C_{\nu}$ at time $t=T$ by 
$x_{\nu} \equiv x_{\nu}(T)$.
Those form the stochastically selected nodes, which define the stochastic basis.

\section{Numerical results}
\label{sec:Results}
\subsection{Quantum mechanics}
\label{sec:QM} 

The Monte Carlo Hamiltonian has been tested on a number of quantum mechanical potential models, by computing the spectrum, wave functions and thermodynamical observables in some temperature window \cite{Jirari:99,Luo:00,Huang:00,Jiang:00}.
Although the results from the Monte Carlo Hamiltonian agree well with the 
exact results, low-dimensional models in Q.M. are not the target of this method. 
The target is rather high-dimensional models or models with a large number of degrees of freedom. The reason for this is based on the same principle 
which applies to numerical integration with Monte Carlo:
The Monte Carlo method is {\it not} the most efficient method when doing low-dimensional integrals. However, it is most efficient when doing integrals in high dimensions (e.g. path integrals).

\subsection{Klein-Gordon model}
\label{sec:KleinGord}

We consider in D=1 a chain of coupled harmonic oscillators, which is equivalent to the Klein-Gordon field on a $1+1$ lattice.
Here we consider a chain of oscillators. 
The model is given by
\bea
\label{eq:Chain}
S &=& \int dt ~ T - V
\nonumber \\
T &=& \sum_{n=1}^{N} \frac{1}{2} m \dot{q}_{n}^{2}
\nonumber \\
V &=& \frac{1}{2} \sum_{n=1}^{N} \Omega^{2}(q_{n} - q_{n+1})^{2} + \Omega_{0}^{2} q_{n}^{2} ~ .
\eea
The parameters have been chosen as
$m=1$, $\Omega=1$, $\Omega_0=2$,
$N_{\rm{osc}}=9$, $a=1$, $T=2$ and $\hbar=1$.
For the adjustable parameter $\sigma$ in the stochastic basis,
we choose $\sigma=\sqrt{\hbar {\rm sinh}(\Omega T)/(m \Omega)}$.
After the stochastic basis with $N_{stoch}=1000$ is generated,
we obtain the matrix elements $M_{n'n}$
Then we compute the eigenvalues and eigenvectors
using the method described in Sect. 
Table \ref{tab.3} gives a comparison between the spectrum
from the effective Hamiltonian with the stochastic basis
and the analytic result for the first 20 states.
They agree very well.
This means that the Gaussian distribution for
the stochastic basis works well.
\begin{table}[hbt]
\caption{Spectrum of Klein-Gordon model on the lattice,
MC Hamiltonian (stochastic basis) versus
exact result.}
\vspace{3mm}
\begin{center}
\begin{tabular}{|c|c|c|}
\hline
$n$ & $E_{n}^{\rm{eff}}$  & $E_{n}^{\rm{exact}}$\\
\hline
   1   & 10.904663192168  &  10.944060480668\\
   2   & 12.956830557334  &  12.944060480668\\
   3   & 12.985023578737  &  13.057803869484\\
   4   & 13.044311582647  &  13.057803869484\\
   5   & 13.299967341242  &  13.321601993380\\
   6   & 13.345480638394  &  13.321601993380\\
   7   & 13.552195133687  &  13.589811791733\\
   8   & 13.585794986361  &  13.589811791733\\
   9   & 13.680136748933  &  13.751084748745\\
   10  &  13.744919087477 &   13.751084748745\\
   11  &  14.984737011385 &   14.944060480668\\
   12  &  15.012353803145 &   15.057803869484\\
   13  &  15.057295761044 &   15.057803869484\\
   14  &  15.108904652020 &   15.171547258300\\
   15  &  15.125356713561 &   15.171547258300\\
   16  &  15.187413290039 &   15.171547258300\\
   17  &  15.308536490102 &   15.321601993380\\
   18  &  15.396255686587 &   15.321601993380\\
   19  &  15.420708031412 &   15.435345382196\\
   20  &  15.432823810789 &   15.435345382196\\
\hline
\end{tabular}
\end{center}
\vspace{0mm}
\label{tab.3}
\end{table}
\\

\noindent We have also computed thermodynamical quantities
such as the partition function $Z$, free energy $F$,
average energy $U=\overline{E}$ and specific heat $C$.
The analytical results are
\begin{eqnarray}
Z(\beta) &=& {\rm Tr} \left( \exp \left( -\beta H \right) \right)
=\prod_{l=1}^{N_{\rm{osc}}} {1 \over 2 {\rm sinh} \left( \beta \hbar
\omega_l/2 \right)},
\nonumber \\
\overline{E}(\beta) &=& {1 \over Z} {\rm Tr} \left(H \exp \left( -\beta H
\right) \right)
= - {\partial \log Z \over \partial \beta}
= \sum_{l=1}^{N_{\rm{osc}}} {\hbar \omega_l \over 2} \coth \left( \beta
\hbar \omega_l/2 \right),
\nonumber \\
C(\beta) &=& k_B {\partial \overline{E} \over \partial {\cal T}}
= -k_B \beta^2 {\partial \overline{E} \over \partial \beta}
= k_B \sum_{l=1}^{N_{\rm{osc}}}
\left( {\beta \hbar \omega_l/2 \over 2 {\rm sinh}
\left( \beta \hbar \omega_l/2 \right)} \right)^2 ,
\end{eqnarray}
where 
\begin{eqnarray}
\omega_l &=& \sqrt{\Omega_0^2+4\Omega^2 sin^2(p_l \Delta x/2)} ~ ,
\nonumber \\
\Delta p &=& 2 \pi / (N_{\rm{osc}} \Delta x) ~ ,
\nonumber \\
x_j &=& [-(N_{\rm{osc}} -1) / 2 + (j-1)]\Delta x ~ ,
\nonumber \\
p_l &=& [-(N_{\rm{osc}} -1) / 2 + (l-1)]\Delta p ~ .
\end{eqnarray}
Here $j$ and $l$ run from 1 to $N_{\rm{osc}}$ (number of oscillators).
$\Delta x=a=1$ is the lattice spacing, $\beta=T/\hbar$, 
the temperature is related to $\beta$ via ${\cal T} = 1/(\beta k_B)$, 
and $k_B$ is the Boltzmann constant. 
\begin{figure}[htb]
\begin{center}
\setlength{\unitlength}{0.240900pt}
\ifx\plotpoint\undefined\newsavebox{\plotpoint}\fi
\sbox{\plotpoint}{\rule[-0.200pt]{0.400pt}{0.400pt}}%
\begin{picture}(1349,809)(0,0)
\font\gnuplot=cmr10 at 10pt
\gnuplot
\sbox{\plotpoint}{\rule[-0.200pt]{0.400pt}{0.400pt}}%
\put(121.0,123.0){\rule[-0.200pt]{4.818pt}{0.400pt}}
\put(101,123){\makebox(0,0)[r]{0}}
\put(1308.0,123.0){\rule[-0.200pt]{4.818pt}{0.400pt}}
\put(121.0,204.0){\rule[-0.200pt]{4.818pt}{0.400pt}}
\put(101,204){\makebox(0,0)[r]{1}}
\put(1308.0,204.0){\rule[-0.200pt]{4.818pt}{0.400pt}}
\put(121.0,285.0){\rule[-0.200pt]{4.818pt}{0.400pt}}
\put(101,285){\makebox(0,0)[r]{2}}
\put(1308.0,285.0){\rule[-0.200pt]{4.818pt}{0.400pt}}
\put(121.0,365.0){\rule[-0.200pt]{4.818pt}{0.400pt}}
\put(101,365){\makebox(0,0)[r]{3}}
\put(1308.0,365.0){\rule[-0.200pt]{4.818pt}{0.400pt}}
\put(121.0,446.0){\rule[-0.200pt]{4.818pt}{0.400pt}}
\put(101,446){\makebox(0,0)[r]{4}}
\put(1308.0,446.0){\rule[-0.200pt]{4.818pt}{0.400pt}}
\put(121.0,527.0){\rule[-0.200pt]{4.818pt}{0.400pt}}
\put(101,527){\makebox(0,0)[r]{5}}
\put(1308.0,527.0){\rule[-0.200pt]{4.818pt}{0.400pt}}
\put(121.0,608.0){\rule[-0.200pt]{4.818pt}{0.400pt}}
\put(101,608){\makebox(0,0)[r]{6}}
\put(1308.0,608.0){\rule[-0.200pt]{4.818pt}{0.400pt}}
\put(121.0,688.0){\rule[-0.200pt]{4.818pt}{0.400pt}}
\put(101,688){\makebox(0,0)[r]{7}}
\put(1308.0,688.0){\rule[-0.200pt]{4.818pt}{0.400pt}}
\put(121.0,769.0){\rule[-0.200pt]{4.818pt}{0.400pt}}
\put(101,769){\makebox(0,0)[r]{8}}
\put(1308.0,769.0){\rule[-0.200pt]{4.818pt}{0.400pt}}
\put(121.0,123.0){\rule[-0.200pt]{0.400pt}{4.818pt}}
\put(121,82){\makebox(0,0){0}}
\put(121.0,749.0){\rule[-0.200pt]{0.400pt}{4.818pt}}
\put(242.0,123.0){\rule[-0.200pt]{0.400pt}{4.818pt}}
\put(242,82){\makebox(0,0){1}}
\put(242.0,749.0){\rule[-0.200pt]{0.400pt}{4.818pt}}
\put(362.0,123.0){\rule[-0.200pt]{0.400pt}{4.818pt}}
\put(362,82){\makebox(0,0){2}}
\put(362.0,749.0){\rule[-0.200pt]{0.400pt}{4.818pt}}
\put(483.0,123.0){\rule[-0.200pt]{0.400pt}{4.818pt}}
\put(483,82){\makebox(0,0){3}}
\put(483.0,749.0){\rule[-0.200pt]{0.400pt}{4.818pt}}
\put(604.0,123.0){\rule[-0.200pt]{0.400pt}{4.818pt}}
\put(604,82){\makebox(0,0){4}}
\put(604.0,749.0){\rule[-0.200pt]{0.400pt}{4.818pt}}
\put(725.0,123.0){\rule[-0.200pt]{0.400pt}{4.818pt}}
\put(725,82){\makebox(0,0){5}}
\put(725.0,749.0){\rule[-0.200pt]{0.400pt}{4.818pt}}
\put(845.0,123.0){\rule[-0.200pt]{0.400pt}{4.818pt}}
\put(845,82){\makebox(0,0){6}}
\put(845.0,749.0){\rule[-0.200pt]{0.400pt}{4.818pt}}
\put(966.0,123.0){\rule[-0.200pt]{0.400pt}{4.818pt}}
\put(966,82){\makebox(0,0){7}}
\put(966.0,749.0){\rule[-0.200pt]{0.400pt}{4.818pt}}
\put(1087.0,123.0){\rule[-0.200pt]{0.400pt}{4.818pt}}
\put(1087,82){\makebox(0,0){8}}
\put(1087.0,749.0){\rule[-0.200pt]{0.400pt}{4.818pt}}
\put(1207.0,123.0){\rule[-0.200pt]{0.400pt}{4.818pt}}
\put(1207,82){\makebox(0,0){9}}
\put(1207.0,749.0){\rule[-0.200pt]{0.400pt}{4.818pt}}
\put(1328.0,123.0){\rule[-0.200pt]{0.400pt}{4.818pt}}
\put(1328,82){\makebox(0,0){10}}
\put(1328.0,749.0){\rule[-0.200pt]{0.400pt}{4.818pt}}
\put(121.0,123.0){\rule[-0.200pt]{290.766pt}{0.400pt}}
\put(1328.0,123.0){\rule[-0.200pt]{0.400pt}{155.621pt}}
\put(121.0,769.0){\rule[-0.200pt]{290.766pt}{0.400pt}}
\put(30,446){\makebox(0,0){$C/k_B$}}
\put(724,21){\makebox(0,0){$\beta$}}
\put(121.0,123.0){\rule[-0.200pt]{0.400pt}{155.621pt}}
\put(1156,729){\makebox(0,0)[r]{MC Hamiltonian}}
\put(181,327){\raisebox{-.8pt}{\makebox(0,0){$\Delta$}}}
\put(242,528){\raisebox{-.8pt}{\makebox(0,0){$\Delta$}}}
\put(302,386){\raisebox{-.8pt}{\makebox(0,0){$\Delta$}}}
\put(362,261){\raisebox{-.8pt}{\makebox(0,0){$\Delta$}}}
\put(423,189){\raisebox{-.8pt}{\makebox(0,0){$\Delta$}}}
\put(483,153){\raisebox{-.8pt}{\makebox(0,0){$\Delta$}}}
\put(543,136){\raisebox{-.8pt}{\makebox(0,0){$\Delta$}}}
\put(604,129){\raisebox{-.8pt}{\makebox(0,0){$\Delta$}}}
\put(664,125){\raisebox{-.8pt}{\makebox(0,0){$\Delta$}}}
\put(725,124){\raisebox{-.8pt}{\makebox(0,0){$\Delta$}}}
\put(785,123){\raisebox{-.8pt}{\makebox(0,0){$\Delta$}}}
\put(845,123){\raisebox{-.8pt}{\makebox(0,0){$\Delta$}}}
\put(906,123){\raisebox{-.8pt}{\makebox(0,0){$\Delta$}}}
\put(966,123){\raisebox{-.8pt}{\makebox(0,0){$\Delta$}}}
\put(1026,123){\raisebox{-.8pt}{\makebox(0,0){$\Delta$}}}
\put(1087,123){\raisebox{-.8pt}{\makebox(0,0){$\Delta$}}}
\put(1147,123){\raisebox{-.8pt}{\makebox(0,0){$\Delta$}}}
\put(1207,123){\raisebox{-.8pt}{\makebox(0,0){$\Delta$}}}
\put(1268,123){\raisebox{-.8pt}{\makebox(0,0){$\Delta$}}}
\put(1328,123){\raisebox{-.8pt}{\makebox(0,0){$\Delta$}}}
\put(1232,729){\raisebox{-.8pt}{\makebox(0,0){$\Delta$}}}
\put(1156,688){\makebox(0,0)[r]{Analytical result}}
\multiput(1176,688)(20.756,0.000){6}{\usebox{\plotpoint}}
\put(1288,688){\usebox{\plotpoint}}
\put(181,766){\usebox{\plotpoint}}
\multiput(181,766)(6.563,-19.690){2}{\usebox{\plotpoint}}
\multiput(193,730)(6.563,-19.690){2}{\usebox{\plotpoint}}
\multiput(205,694)(6.065,-19.850){2}{\usebox{\plotpoint}}
\multiput(216,658)(6.563,-19.690){2}{\usebox{\plotpoint}}
\multiput(228,622)(5.915,-19.895){2}{\usebox{\plotpoint}}
\multiput(239,585)(6.403,-19.743){2}{\usebox{\plotpoint}}
\put(256.72,528.75){\usebox{\plotpoint}}
\multiput(262,511)(6.403,-19.743){2}{\usebox{\plotpoint}}
\multiput(274,474)(6.732,-19.634){2}{\usebox{\plotpoint}}
\multiput(286,439)(6.563,-19.690){2}{\usebox{\plotpoint}}
\put(302.81,391.00){\usebox{\plotpoint}}
\multiput(309,375)(7.589,-19.318){2}{\usebox{\plotpoint}}
\put(326.30,333.35){\usebox{\plotpoint}}
\put(335.31,314.66){\usebox{\plotpoint}}
\multiput(344,298)(9.631,-18.386){2}{\usebox{\plotpoint}}
\put(365.56,260.29){\usebox{\plotpoint}}
\put(376.81,242.84){\usebox{\plotpoint}}
\put(389.59,226.51){\usebox{\plotpoint}}
\put(402.63,210.37){\usebox{\plotpoint}}
\put(417.67,196.10){\usebox{\plotpoint}}
\put(433.68,182.90){\usebox{\plotpoint}}
\put(450.82,171.21){\usebox{\plotpoint}}
\put(468.89,161.05){\usebox{\plotpoint}}
\put(487.83,152.57){\usebox{\plotpoint}}
\put(507.37,145.63){\usebox{\plotpoint}}
\put(527.45,140.39){\usebox{\plotpoint}}
\put(547.74,136.06){\usebox{\plotpoint}}
\put(568.25,133.23){\usebox{\plotpoint}}
\put(588.83,130.67){\usebox{\plotpoint}}
\put(609.43,129.00){\usebox{\plotpoint}}
\put(630.11,127.24){\usebox{\plotpoint}}
\put(650.82,126.00){\usebox{\plotpoint}}
\put(671.53,125.00){\usebox{\plotpoint}}
\put(692.28,124.89){\usebox{\plotpoint}}
\put(713.00,124.00){\usebox{\plotpoint}}
\put(733.75,124.00){\usebox{\plotpoint}}
\put(754.51,124.00){\usebox{\plotpoint}}
\put(775.25,123.73){\usebox{\plotpoint}}
\put(795.98,123.00){\usebox{\plotpoint}}
\put(816.73,123.00){\usebox{\plotpoint}}
\put(837.49,123.00){\usebox{\plotpoint}}
\put(858.25,123.00){\usebox{\plotpoint}}
\put(879.00,123.00){\usebox{\plotpoint}}
\put(899.76,123.00){\usebox{\plotpoint}}
\put(920.51,123.00){\usebox{\plotpoint}}
\put(941.27,123.00){\usebox{\plotpoint}}
\put(962.02,123.00){\usebox{\plotpoint}}
\put(982.78,123.00){\usebox{\plotpoint}}
\put(1003.53,123.00){\usebox{\plotpoint}}
\put(1024.29,123.00){\usebox{\plotpoint}}
\put(1045.05,123.00){\usebox{\plotpoint}}
\put(1065.80,123.00){\usebox{\plotpoint}}
\put(1086.56,123.00){\usebox{\plotpoint}}
\put(1107.31,123.00){\usebox{\plotpoint}}
\put(1128.07,123.00){\usebox{\plotpoint}}
\put(1148.82,123.00){\usebox{\plotpoint}}
\put(1169.58,123.00){\usebox{\plotpoint}}
\put(1190.33,123.00){\usebox{\plotpoint}}
\put(1211.09,123.00){\usebox{\plotpoint}}
\put(1231.84,123.00){\usebox{\plotpoint}}
\put(1252.60,123.00){\usebox{\plotpoint}}
\put(1273.36,123.00){\usebox{\plotpoint}}
\put(1294.11,123.00){\usebox{\plotpoint}}
\put(1314.87,123.00){\usebox{\plotpoint}}
\put(1328,123){\usebox{\plotpoint}}
\end{picture}
\end{center}
\caption{Specific heat ($C / k_B$) of the Klein-Gordon model on a 1+1
dimensional lattice.}
\label{fig.3}
\end{figure}
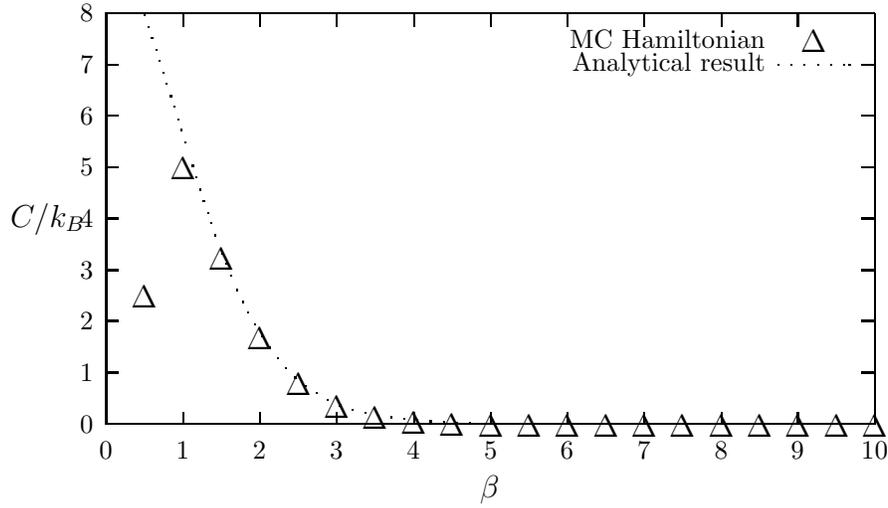
Since we have approximated $H$ by $H_{\rm{eff}}$,
we can express those thermodynamical observables
via the eigenvalues of the effective Hamiltonian
\begin{eqnarray}
Z^{\rm{eff}}(\beta) &=& \sum_{n=1}^{N}e^{-\beta E_{n}^{\rm{eff}}},
\nonumber \\
\overline{E}^{\rm{eff}}(\beta) &=& \sum_{n=1}^{N}
{E_{n}^{\rm{eff}} {\rm e}^{-\beta E_{n}^{\rm{eff}}} \over
Z^{\rm{eff}}(\beta)},
\nonumber \\
C^{\rm{eff}}(\beta) &=& k_B{\beta}^2\left(\sum_{n=1}^{N}
{(E_{n}^{\rm{eff}})^2e^{-\beta E_{n}^{\rm{eff}}} \over
Z^{\rm{eff}}(\beta)}-\left(\overline{E}^{\rm{eff}}(\beta)\right)^2 \right).
\end{eqnarray}
Since this is a static system, the eigenvalues should in principle not vary
with $\beta$.
This is observed numerically within statistical errors when $\beta$ and $N_{stoch}$ are not too small.
Using this assumption and
the spectrum at $\beta=2$, we can compute thermodynamical quantities for
other values of $\beta$.
The specific heat
as a function of $\beta$ is shown in Fig. [1].
Again, the results from the MC Hamiltonian are in good agreement
with the analytical ones when $\beta > 1$.
Preliminary results for larger $N_{{\rm osc}}$ indicate that it is not
necessary to increase $N_{stoch}$ accordingly.
This property is very important for an application of the algorithm
to many-body systems and QFT.

\section{Summary}

In this paper, we have extended the effective Hamiltonian method
with a stochastic basis to QFT, and taken the Klein-Gordon model as an example.
The results are very encouraging. We believe that the application
of the algorithm to more complicated systems will be very interesting. It will allow a non-perturbative investigation of physics beyond the ground state.

\vspace{1.0cm}

\noindent {\bf Acknowledgements} \\ 
H.K. and K.M. are grateful for support by NSERC Canada. 
X.Q.L. has been supported by NSF for Distinguished Young Scientists of China, by Guangdong Provincial NSF and by the Ministry of Education of China.

\end{document}